# Cotunneling effects in GaAs vertical double quantum dots

A.O. Badrutdinov[1,2], S. M. Huang[2], K. Kono[2], K. Ono[2] and D. A. Tayurskii[1]

[1]*Physics Department, Kazan Federal University, 420008, Kazan, Russia*
[2]*Low Temperature Physics Laboratory, Advanced Science Institute, RIKEN, Wako, 351-0198, Japan*

Semiconductor quantum dots have been a subject of intensive investigation over last two decades. One reason is the proposal [1] to use spin state of electron in a quantum dot as a quantum bit of information (qubit). Another is that quantum dots are systems where a lot of interesting phenomena is observed. In fact, physics of quantum dots has a lot of parallels with atomic physics, and in literature quantum dots are called artificial atoms [2]. At the moment, plenty of work has been already done, to understand physics of quantum dots and to develop methods to adopt quantum dots for practical applications [2, 3, 4]. However, still there are a lot of open questions, and the subject attracts a lot of attention.

In the present paper we report observation of Coulomb blockade lifting in GaAs vertical double quantum dot caused by cotunneling processes. Figure 1a shows schematic view of our sample. The device is a sub-micron pillar structure (~500 nm diameter) containing two 12 nm thick $In_{0.05}Ga_{0.95}As$ quantum wells, separated by pure GaAs potential barriers. The pillar is surrounded with a gate electrode, which allows to tune the potential in both quantum wells. Quantum wells can exchange electrons with reservoirs made of n-doped GaAs. The details of sample fabrication process are described in [5]. We note that the height of the tunnel barriers in the investigated device is relatively low, compared to the usual case of AlGaAs barriers. This results in high tunneling rates through the double dot.

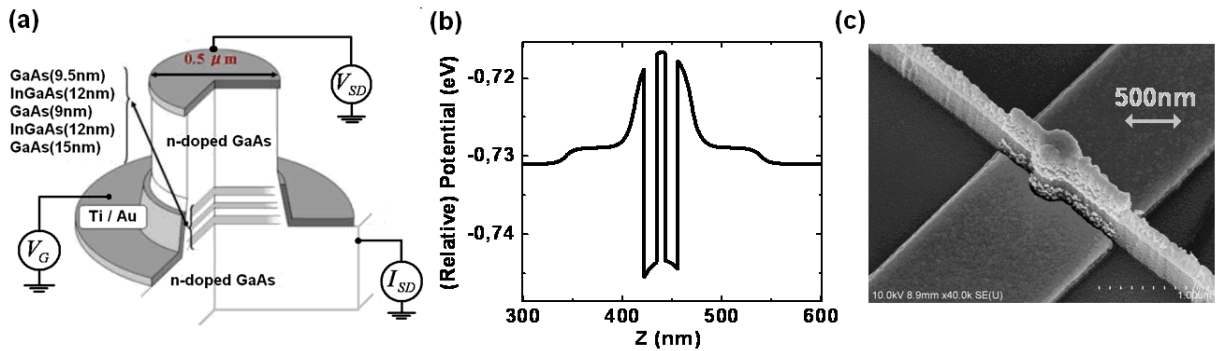

Figure 1. (a) Schematic view of investigated double quantum dot. Dots are defined in $In_{0.05}Ga_{0.95}As$ quantum wells. Potential barriers are formed by pure GaAs layers. (b) Calculated potential profile of original heterostructure, from which the quantum dot was fabricated, in vertical direction. Small asymmetry relative to center barrier reflects the effect of applied source-drain bias. (c) Scanning electron microscope image of a device identical to investigated one.

Figure 2 shows a color plot of differential conductance $dI/dV_{SD}$ of our sample, as a function of source-drain and gate voltages, measured at dilution fridge temperature of 10 mK and zero magnetic field. Dark blue lines observed on the plot correspond to step like change of current through the double dot, which reflects the discrete energy spectrum. Analysis of differential conductance diagram is performed using conventional constant interaction model [3, 4], where the expressions for electrochemical potentials of two dots are written as

$$\mu_1(N_1, N_2) = (N_1 - \frac{1}{2})E_{C1} + N_2 E_{Cm} - \alpha_1 V_S - \beta_1 V_D - \gamma_1 V_G + E_{N1} + \delta$$

$$\mu_2(N_1, N_2) = (N_2 - \frac{1}{2})E_{C2} + N_1 E_{Cm} - \alpha_2 V_D - \beta_2 V_S - \gamma_2 V_G + E_{N2}$$

We derive the values of intradot Coulomb energy $E_{C1}$, $E_{C2}$ = 4.5 $meV$, interdot Coulomb energy $E_{Cm}$ = 2 $meV$, and the lowest energy level difference between two dots $\delta$ = 3.5 $meV$. The coefficients of electrostatic coupling between the dots and source, drain and gate electrodes are determined to be $\alpha_1 = \beta_2 = 0.24e$, $\alpha_2 = \beta_1 = 0.76e$ $\gamma_1 = \gamma_2 = 0.07e$, where $e$ is elementary charge modulus.

One feature which can not be explained in terms of constant interaction model is observation of finite differential conductance in the areas of diagram where dot should be under Coulomb blockade (on the diagram this looks like blue triangles in Coulomb diamond areas). We explain this in terms of cotunneling mechanism, which has been described previously for the case of a single quantum dot [6]. We focus on (0,2) charge state, where the corresponding area is most clear. Along the line separating areas A and B on the schematic diagram of figure 2, condition $\mu_S - \mu_D = \mu_1(1,1) - \mu_2(0,2)$ is satisfied. In the area A, where $\mu_S - \mu_D > \mu_1(1,1) - \mu_2(0,2)$, virtual process of simultaneous tunneling from dot 2 to drain and from source to dot 1, as shown on figure 2, is energetically possible. This leads to finite current through the dot, as observed on the differential conductance diagram. In area B, where $\mu_S - \mu_D < \mu_1(1,1) - \mu_2(0,2)$, energy conservation law is not satisfied for this process, and dot is under Coulomb blockade.

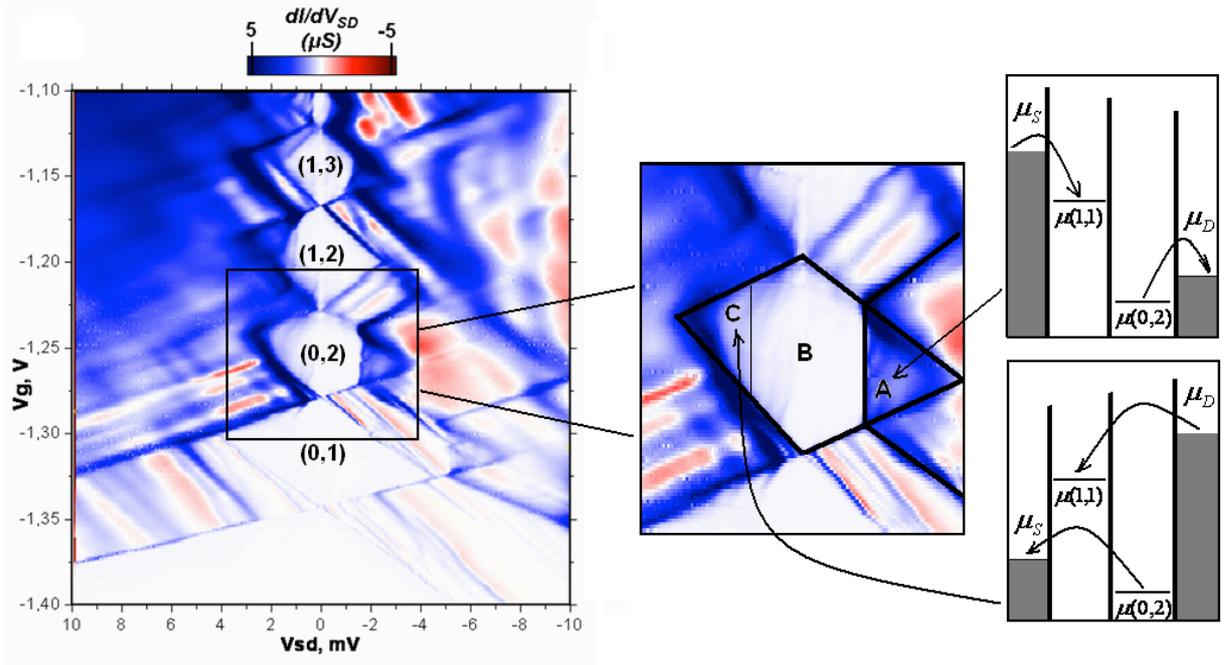

One noticeable difference of cotunneling in double dot compared to single dot case is the asymmetry under sign of applied source-drain bias. For (0,2) charge state we see that there is cotunneling area at negative source-drain voltage, but quite usual shape of Coulomb diamond is observed at positive source-drain voltage. In contrast, cotunneling observed in single dot was symmetric under applied voltage sign [6]. This can be explained in the following way. Figure 2 shows the relative position of electrochemical potentials of involved states, in case of double dot under opposite sign of applied bias voltage (area C). In this case each dot forms a potential barrier for electron tunneling from (to) another dot. It makes probability of such transition very low and mechanism becomes totally inefficient. This is not the case for single dot, where the effective barriers are symmetric for cotunneling processes under both signs of applied voltage.

Figure 2. (left) Measured differential conductance $dI/dV_{SD}$, as a function of $V_{SD}$ and $V_G$, of the investigated double quantum dot. The numbers ($N_1$, $N_2$) indicate stable charge configuration, when dot 1 (2) contains $N_1$ ($N_2$) electrons. (right) Part of measured differential conductance diagram showing Coulomb blockade region where dot is in (0,2) charge state. Schematic diagrams show the relative positions of electrochemical potentials, in the areas of Coulomb diamond where $|\mu_S - \mu_D| > |\mu_1(1,1) - \mu_2(0,2)|$, for both signs of $V_{SD}$.

An interesting consequence of asymmetry discussed above is correlation between position of cotunneling area and order of dot filling with electrons. Our double dot is filled in sequence (0,1) – (0,2) – (1,2) – (1,3) – (…). In case of $1^{st}$, $2^{nd}$ and $4^{th}$ Coulomb diamonds the last electron which entered the dot resides in dot 2, and we observe cotunneling induced current at negative source drain bias. In contrast, in case of $3^{rd}$ Coulomb diamond last electron resides in dot 1, and relative position of electrochemical potentials necessary for cotunneling happens at positive bias. We also note, that in case of $1^{st}$ diamond cotunneling triangle takes place not in the Coulomb blockade area directly, but in the nearby area of second order tunneling. This is the result of initial potential offset between dot 1 and dot 2. In this cotunneling area (0,2) charge state is also in the bias window, so the current results from both cotunneling mechanism involving (1,0) and (0,1) charge states, and from second order tunneling involving (0,2) charge state.

In conclusion, we observed lifting of Coulomb blockade in GaAs vertical double quantum dot with low potential barriers, induced by cotunneling mechanisms. Several distinct features were observed, compared to single dot case, and appropriate explanation for them was given.

A. B. and D. T. are supported in part by the Ministry of Education and Science of the Russian Federation (contract N 02.740.11.0797).

References:

1. D. Loss and D. P. DiVincenzo, 1998, Phys. Rev. A **57**, 120.
2. L. P. Kouwenhoven et al., 2001, Rep. Prog. Phys. **64**, 701-736.
3. W. G. van der Wiel et al., 2003, Rev. Mod. Phys. **75**, 1.
4. R. Hanson et al., 2007, Rev. Mod. Phys. **79**, 1217.
5. D. G. Austing et al., 1996, Semicond. Sci. Technol. **11**, 388.
6. S. De Franceschi et al., 2001, Phys. Rev. Lett. **86**, 878.